\documentclass{aa}
\usepackage{natbib}
\usepackage{graphicx}
\usepackage{txfonts}
\begin{document}
   \title{A compact flare eclipsed in the corona of SV Cam}

   \author{J. Sanz-Forcada\inst{1}
          \and
          F. Favata\inst{1}
          \and
          G. Micela\inst{2}
          }

   \offprints{J. Sanz-Forcada, \email{jsanz@rssd.esa.int}}

   \institute{Astrophysics Division -- Research and Science Support Department
  	of ESA, ESTEC, Postbus 299, NL-2200 AG Noordwijk, The Netherlands\\
              \email{jsanz@rssd.esa.int,ffavata@rssd.esa.int}
         \and
             INAF - Osservatorio Astronomico di Palermo,
             Piazza del Parlamento 1, I-90134 Palermo, Italy
             \email{giusi@astropa.unipa.it}
             }

   \date{Received / Accepted }

   \abstract{The eclipsing active binary SV Cam (G0V/K6V, $P_{\rm
   orb}$=0.593071 d) was observed with XMM-{\em Newton}
   during two campaigns in 2001 and 2003. 
   No eclipses in the quiescent emission are clearly identified, but 
   a flare was eclipsed during the 2001 campaign, allowing us to strongly
   constrain, from purely geometrical considerations, 
   the position and size of the event: the flare is compact 
   and it is formed at a latitude below 65$\degr$. The
   size, temperature and Emission Measure of the flare imply an
   electron density of $\log n_{\rm e}$ (cm$^{-3}$)$\sim$10.6--13.3
   and a magnetic field of $\sim$65--1400~G in order to confine the
   plasma, consistent with the measurements that are obtained
   from density-sensitive line ratios in other similar active stars. 
   Average emission seems to come
   from either extended or polar regions because of lack of eclipses. 
   The Emission Measure
   Distribution, coronal abundances and characteristics of
   variability are very similar to other active stars such as AB~Dor (K1V).
   }

   \maketitle
%

\section{Introduction}
The analysis of stellar coronae is commonly based on several aspects
such as thermal structure, coronal abundances, stellar flares and
analysis of eclipses. 
Since the first high resolution spectra in X-rays and
XUV it has become evident that low activity stars, such as the
Sun, have a thermal structure substantially different from that of
active stars. Active stars, such as AB~Dor or Capella are dominated by
material at $T\sim 10$~MK, much hotter than the solar corona ($T \sim
2$~MK), and display a substantially larger number of flares
\citep[][ and references
therein]{dup93,sch95,bow00,survey,huen01,mag04,fav03}. 
The thermal
structure found in active stars 
is normally interpreted as the combination of different
families of coronal loops \citep[e.g.][]{gri98,jar02}, 
and it is reminiscent of the thermal structure found in some solar flares.
The analysis of eclipses and rotational modulation raised the question
on whether these loops with peak
temperature around $\log T$(K)$\sim$6.9 are linked to a given
coronal region such as the stellar poles \citep{bri98,schm99}, where
photospheric spots are usually detected. 
One of the ways to get further
insight in the knowledge of the coronae of active stars is to observe
eclipsing binaries containing active stars, that may allow us to
estimate the size and geometrical position of the emitting regions and
to derive other characteristics.
Coronal abundances of active
stars also tend to show a rather low abundance of low-FIP elements
when they are compared to the solar photosphere, although comparison
with their own photospheric values may result in a composition
consistent with that of the photosphere of the star \citep{sanz04},
this still being an open debate.

The eclipsing binary SV Cam (HD 44982) is a well-studied spectroscopic
RS~CVn system at a distance of 85~pc \citep{hippa}, 
formed by a spectroscopic binary
\citep[G0V/K6V,][]{leh02} and the likely presence of a third (less
massive) body in a wide orbit \citep{alb01}. 
SV Cam has a short orbital period \citep[$P_{\rm
orb}$=0.593071 d,][]{poj98}, with locked rotation that causes 
enhanced stellar activity in the system.
SV Cam has been detected
by ROSAT \citep{hem97}, but the low statistics of the
observation did not allow us to clearly identify the presence of eclipses
in the X-ray band. 
Optical observations reveal the presence of spots in the primary star,
but the low luminosity of the secondary ($\sim$15\% of the total) does
not allow us to identify such features on the cooler object 
\citep[][ and references therein]{hem97,kju02,zbo03}. 
The H$\alpha$ and Ca IR triplet profiles
are filled in due to chromospheric activity in the secondary star
\citep{mon95,poj98,kju02}.
Recently, \citet{jef05} found large polar spots in the photosphere of
the primary star using Doppler imaging techniques, as well as the
presence of smaller spots on the rest of the stellar surface (as is
typical of active stars).

The outline of this paper is as follows: Sect. 2 describes the
observations along with a brief description of the analysis
techniques; the results are explained in Sect. 3, followed by a
discussion, and Sect. 4 presents the conclusions.

\begin{figure}
   \centering
   \includegraphics[angle=90,width=0.45\textwidth]{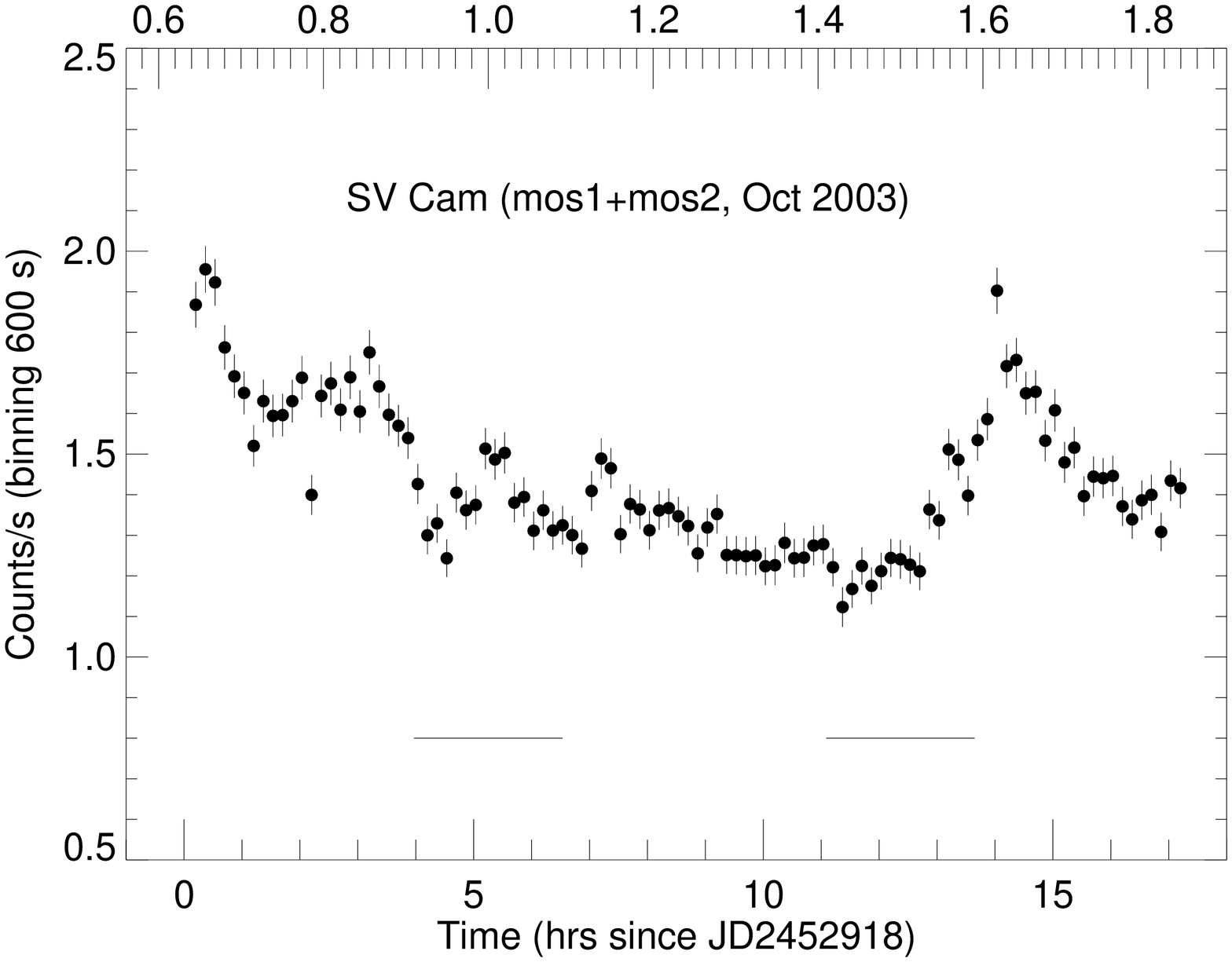}
   \includegraphics[angle=90,width=0.45\textwidth]{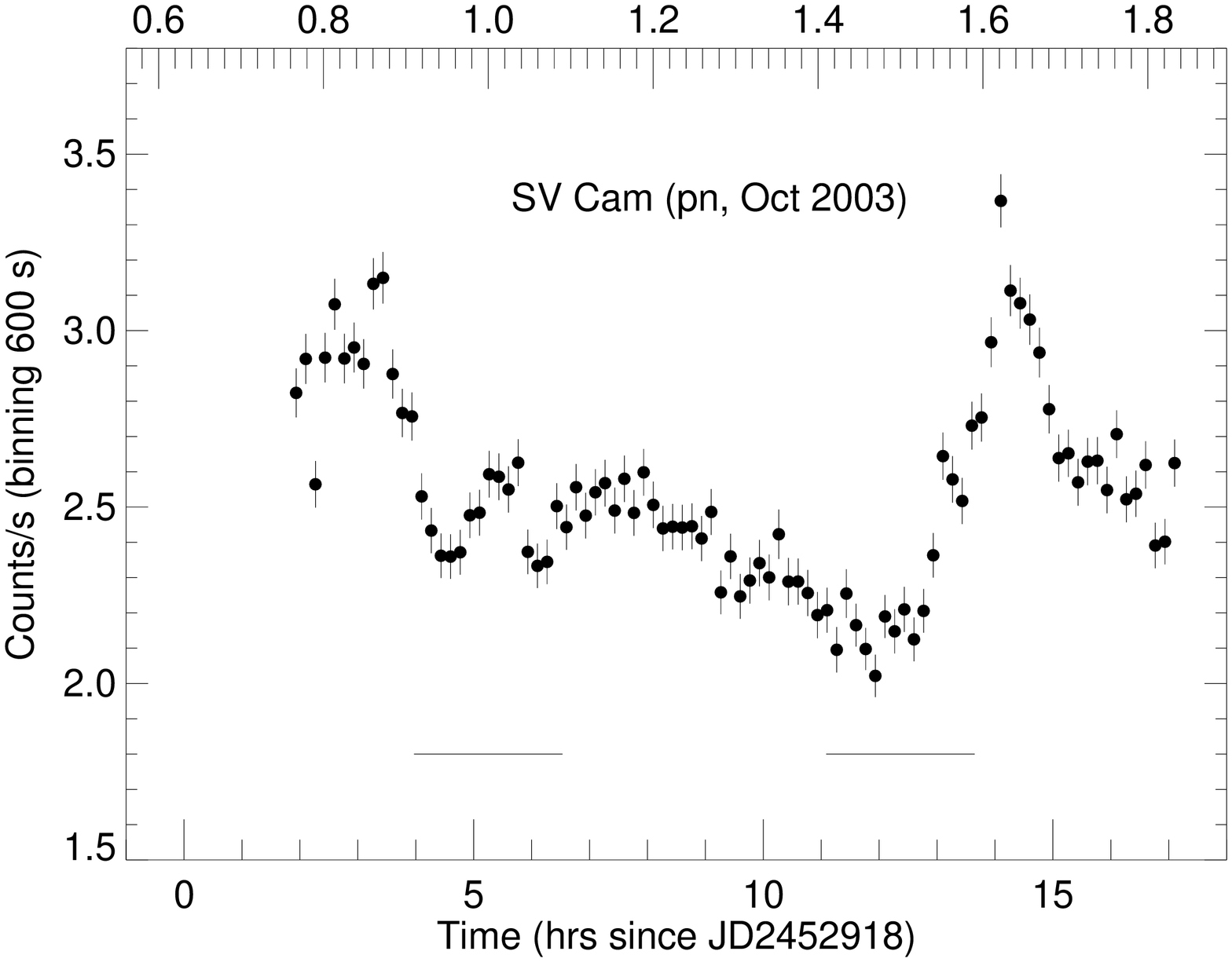}
   \caption{SV Cam light curves (MOS1+MOS2 and PN respectively) 
   in the 2003 campaign, with 1-$\sigma$ error
   bars. The upper axis reports the orbital phase, with $T_0$=HJD
   2449350.3037 
   corresponding to primary (G0V) star located behind the secondary
   (K6V) star, and  $P_{\rm orb}$=0.593071~d \citep{poj98}.
   Solid lines mark the times when photospheric eclipses take place.}
   \label{pnlc}
\end{figure}
\begin{figure}
   \centering
   \includegraphics[angle=90,width=0.45\textwidth]{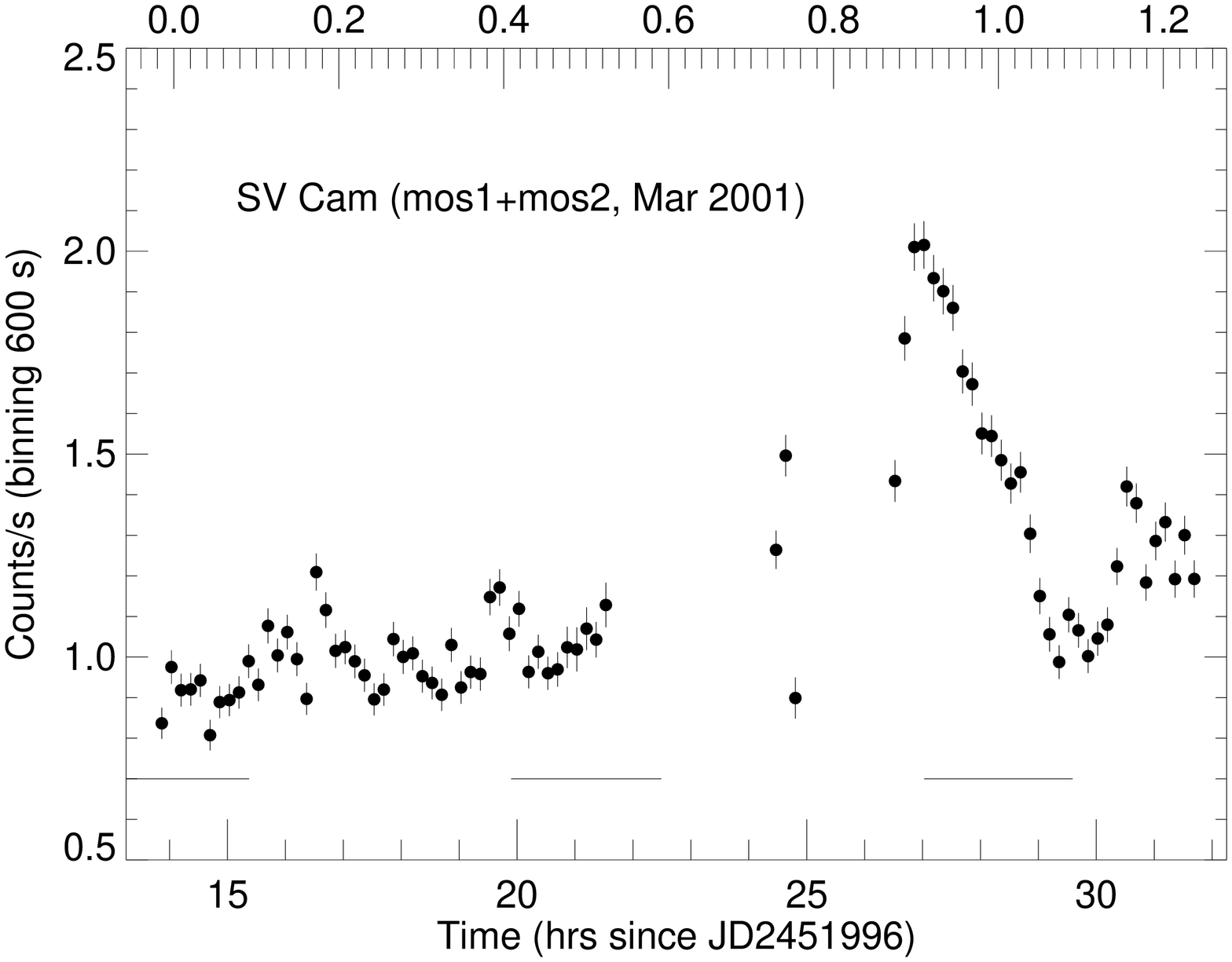}
   \includegraphics[angle=90,width=0.45\textwidth]{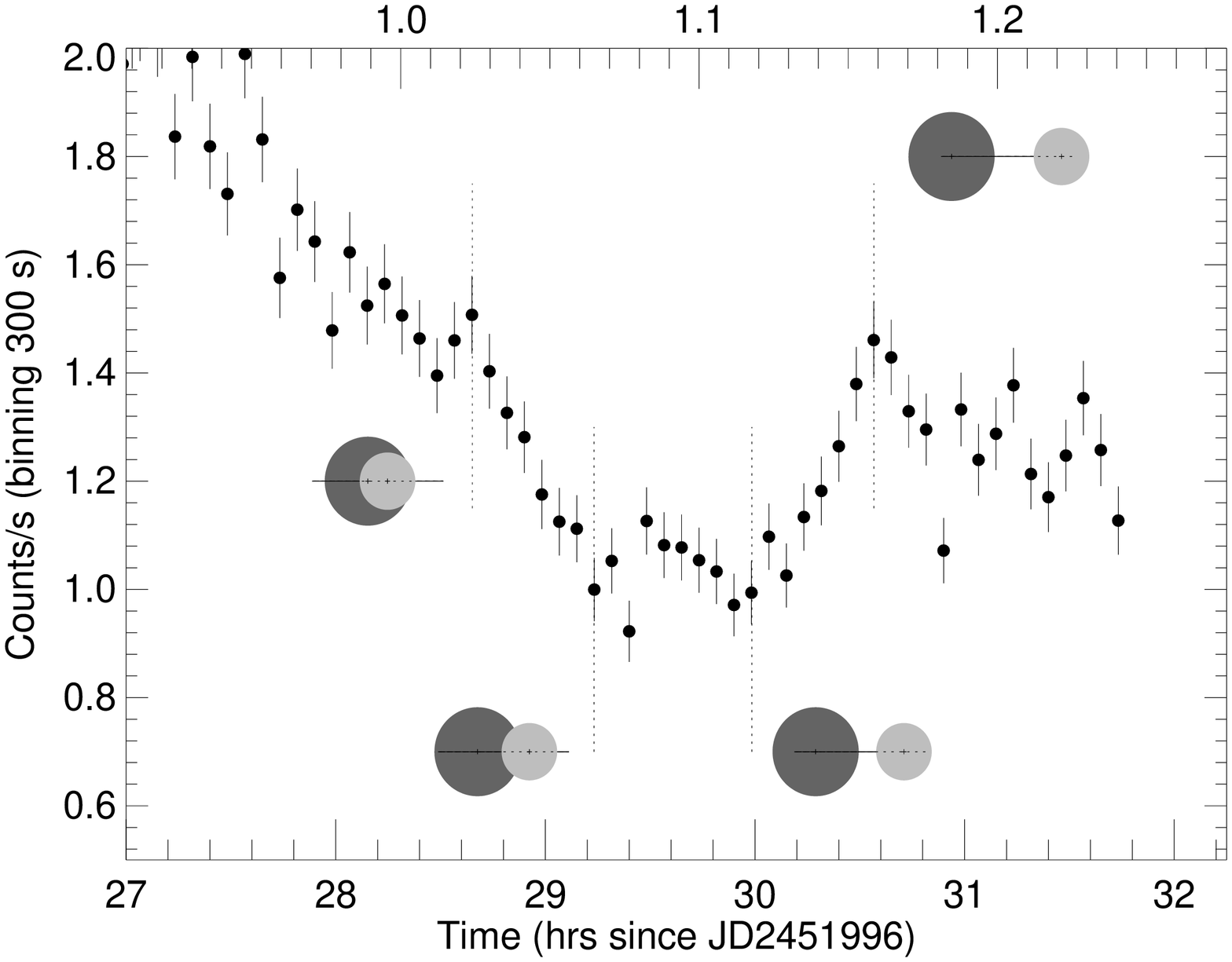}
   \caption{{\it Upper panel}: Same as Fig.~\ref{pnlc}, but for the 2001
   campaign (only MOS data are available). An eclipse in the flare,
   around $\phi \sim$1.1, is displayed in the {\it lower panel} 
   with a bin size of 300~s. 
   Dotted lines indicate the phases used for the four
   contacts. The relative positions of the stars during the contacts
   are also indicated.}
   \label{fig:eclipsezoom}
\end{figure}

\begin{figure*}
   \centering
   \includegraphics[angle=270,width=0.85\textwidth]{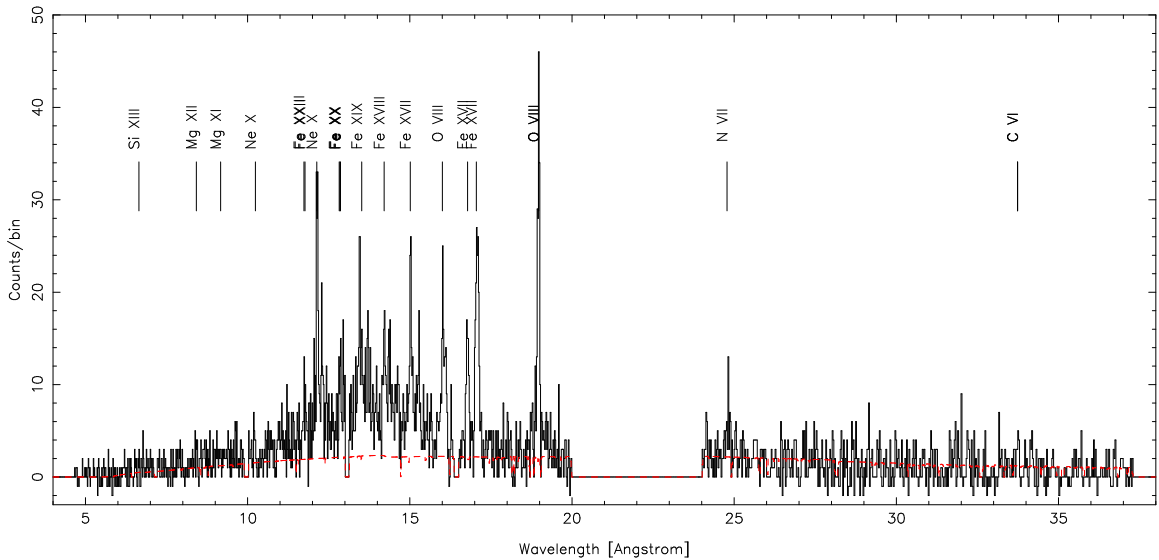}
   \caption{RGS 2 spectrum of SV~Cam in October 2003. The dashed 
 line represents the continuum predicted by the derived EMD. A false continuum
 is created by the extended instrumental line profiles.}  
   \label{fig:rgs}
\end{figure*}

%
\section{Observations}

SV Cam was observed with XMM-\emph{Newton} (P.I. F.~Favata) in March
2001 (rev. \#238) for 65~ks, 
but due to technical problems and high background part
of the data were not usable, in particular the PN data. 
The target was observed again in October
2003 (rev. \#700) for 67~ks.
XMM-\emph{Newton} carries out simultaneous
observation with the EPIC (European Imaging Photon Camera) PN and MOS
detectors (sensitivity range 0.15--15 keV and 0.2--10~keV
respectively), and with the RGS \citep[Reflection Grating
Spectrometer,][]{denher01} ($\lambda\lambda\sim$6--38~\AA,
$\lambda$/$\Delta\lambda\sim$100--500).  The data have been
reduced employing the standard SAS (Science Analysis Software) version
6.1.0 package, removing in the RGS spectra the time intervals when
the background was higher than 0.7 cts/s in CCD \#9, to
ensure a ``clean'' spectrum.  Light curves were obtained by selecting
a circle centered on the source in the EPIC-PN and EPIC-MOS images,
and subtracting the background count rate taken from a nearby area 
(Figs.~\ref{pnlc}, \ref{fig:eclipsezoom}). In the 2001 campaign the
EPIC data were split 
in three intervals due to high flaring background. In the
PN, the third interval suffered from technical problems and therefore
we display only the combined MOS1+MOS2 light curve (data from both
detectors were obtained separately and then combined in the light
curve after fixing the time shift present between MOS 1 and 2).
Mid-resolution spectra were obtained with EPIC detectors
in the range 0.3--10~keV, once the 
intervals with high background level were excluded.
The average X-ray luminosity ($L_{\rm X}$) was calculated from
the PN spectra of the two campaigns (the whole interval was used for
2003, and only the first interval
for 2001), resulting in values of $L_{\rm
X}=1.7\times 10^{30}$ and $L_{\rm X}=2.4\times 10^{30}$~erg~s$^{-1}$ 
in the range
6--20~\AA\ (0.62--2.1 keV) for 2001 and 2003 respectively, and  $L_{\rm
X}=2.6\times 10^{30}$~erg~s$^{-1}$ and $L_{\rm X}=3.9\times
10^{30}$~erg~s$^{-1}$ in the range 
5--100~\AA\ (0.12--2.4~keV), the range covered by ROSAT/PSPC. This
indicates an increase in the X-ray flux by $\ga$50~\%  between the
two observations, with $L_{\rm X}/L_{\rm bol}=5.6\times 10^{-4}$ and
$L_{\rm X}/L_{\rm bol}=8.5\times 10^{-4}$ calculated over the
5--100~\AA\ range.

The rich RGS spectrum taken in 2003 (Fig.~\ref{fig:rgs})
has been used to calculate the thermal
structure (the Emission Measure Distribution, EMD) and coronal
abundances of SV Cam, following the method described in
\citet{sanz03}: line fluxes are measured in the RGS spectrum, and a
trial EMD and a set of coronal abundances are combined with an
atomic plasma model in order to compare 
observed and predicted line fluxes. The EMD and abundances are
then corrected 
to get the best approximation between the two sets of line fluxes
values, and an iterative process is carried out in order to better determine
the continuum position that is employed to measure the line
fluxes in RGS. The determination of abundances during the process is
made as follows: only Fe lines are used initially to determine the
EMD; then the abundance
of Ne is fixed in the temperature domain where the EMD is known, and the EMD
is extended 
afterwards to the rest of the temperature domain covered by Ne; 
the rest of elements are added sequentially following the same process.
This assumes that the relative elemental abundances are constant
throughout the corona.
Such an approach has shown good consistency between results
of the different detectors capable of achieving high-spectral resolution of
stellar coronae \citep{sanz03}.  In this analysis we have measured the
line fluxes (Table~\ref{tab:fluxes}) 
using the {\em Interactive Spectral Interpretation System}
\citep[ISIS,][]{isis} software package, through convolution of the
spectral response. The measured line fluxes were then corrected for
interstellar medium absorption (ISM) using a value of $\log~N_{\rm
H}({\rm cm}^{-2})\sim 19.7$ (although such a correction has very little
influence on the result).
The EMD was reconstructed
using a step in temperature of 0.1 dex, the same employed in the
Astrophysical Plasma Emission Database \citep[APED v1.3.1,][]{aped},
the plasma atomic model used in our analysis.

\section{Results and discussion}

\subsection{Light curves}
\begin{figure}
   \centering
   \includegraphics[angle=90,width=0.45\textwidth]{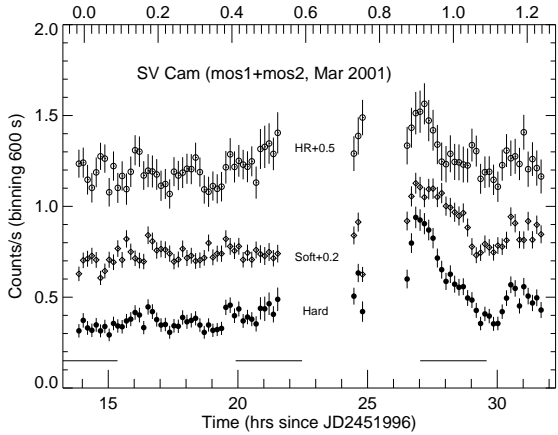}
   \caption{Hardness ratio ({\em HR}), defined as {\em HR}={\em
   Hard}/{\em Soft}, 
   where {\it Hard} and {\it Soft} are
   the fluxes in the bands 1--10~keV and 0.3--1~keV respectively. Note
   the drop in the {\em HR} during the eclipse, at $\phi \sim 1.1$.}  
   \label{fig:hr}
\end{figure}

The light curves of SV~Cam reveal a high level of variability in the
corona of this system, with flares clearly identified in both
campaigns (Figs.~\ref{pnlc}, \ref{fig:eclipsezoom}).
In 2001 a drop in counts occurred while a flare was
decaying, reaching the quiescent levels at phase $\phi \sim 1.1$, and
then returning to the flaring levels. We interpret this decay as an
eclipse of the flaring region, with the totality of this region being
occulted. The light curve of the hardness ratio,
defined as {\em HR}={\em Hard}/{\em Soft} where {\em Hard} and {\em
Soft} are the fluxes in the bands 1--10~keV
and 0.3--1~keV respectively
(Fig.~\ref{fig:hr}) is sensitive to the changes in temperature. The
flare is well identified by the rise in temperature, with a subsequent 
decay. During the eclipse, at $\phi \sim 1.1$ the temperature
drops to the level previous to the flare. 
The existence of two flares could also explain this behaviour,
but two further arguments strengthen the hypothesis of an 
eclipse taking place: (i) the decay of the first peak changes suddenly to
a much steeper one (at $\phi\sim$1.02); although some cases with two decays have been
reported in the literature, the second decay is always less steep than
the first, while in the case of SV~Cam we observe a steeper decay,
something not observed to date;  (ii) the
light curve shows a strong 
symmetry between the points that we interpret as responsible for the 4
contacts, a symmetry much more likely in an eclipsing phenomenon.
Therefore, although we cannot completely rule out a
sequence of two flares, we consider the hypothesis of an eclipse
as the most likely.
We can use this eclipse to derive
some information of the flare, as we will explain in
detail below.  

In the 2003 campaign (Fig.~\ref{pnlc}) the observation appears to take
place during the 
decay of a flare, with a new flare developing at $\phi \sim
1.6$. We find no evidence of an eclipse in the light curve.
Therefore,
from the two campaigns we can conclude that most of the coronal 
emission is not being occulted during eclipses, and therefore the
non-flaring emission is either very extended or it originates at high
latitudes on the primary. 

\begin{figure}
   \centering
   \includegraphics[width=0.49\textwidth]{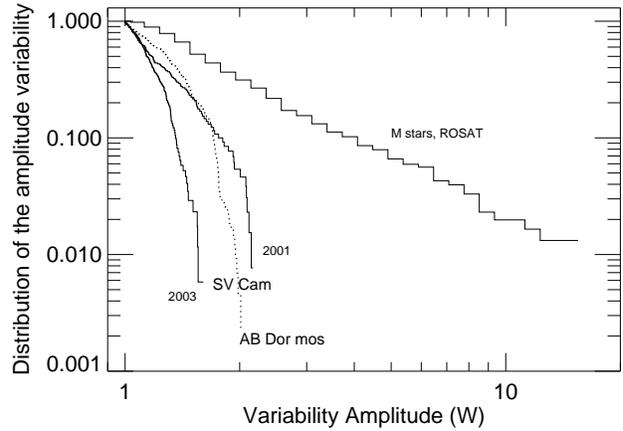}
   \caption{Normalized cumulative distribution of the amplitude
   ($W=L_{\rm X}/L_{\rm min}$) variability in SV Cam in the two
   observing campaigns, as measured in EPIC MOS1+MOS2 count
   rates. Also displayed are the ROSAT/PSPC ($\Delta E$=0.12--2~keV)
   observations of a sample of M stars \citep{mar00} and the MOS
   1+MOS 2 amplitude variability in observations of AB~Dor (see text).}
   \label{amplitude}.
\end{figure}

Regarding the variability of the source, we have made an analysis
of the amplitude variability, defining
amplitude as $W=L_{\rm X}$/$L_{\rm min}$, where $L_{\rm min}$ is the
assumed quiescent value of the light curves (corresponding to
the mode of the count rate distribution), following
\citet{paperadleo}. The amplitude variability curve, shown in
Fig.~\ref{amplitude} tells us how activity ``behaves'' in a star or
group of stars. In Fig.~\ref{amplitude} we compare the case of the two
observations of SV~Cam (each with its own $L_{\rm min}$) with 
the K1V active star AB~Dor (Sanz-Forcada et
al., in preparation) and a group of M
stars \citep{mar00}. The observations of AB~Dor were made with
same instrument (EPIC MOS~1+MOS~2), during XMM-{\em Newton} 
revolutions \#162,
205, 266, 338, 429 and 462. The M stars were observed with    
ROSAT in the range $\Delta E$=0.12--2~keV. 
Although the spectral range and time scale of the 
observations is different, a contrast between the
variability in active M stars and that from the active G and K dwarfs
SV~Cam and AB~Dor is quite evident. 
Results from EUVE ($\Delta E$=0.07--0.15~keV) for
the dM star AD~Leo have shown very similar results to those of ROSAT
\citep{paperadleo}.  Among the two observations of SV Cam, in 2003
there is a higher level of emission, but a lower level of
variability than in 2001. The latter is also remarkably similar to
the average behavior of AB~Dor.

The whole view of the variability in SV~Cam shows an active star with
a quiescent level rather difficult to identify, and with flaring
activity going on in the two campaigns. No eclipses out
of flares could be identified, but an eclipse during a flare
has been observed, and it will be the subject of further analysis
below. These light curves indicate that the 
``quiescent'' coronal emission comes either from an
extended region, or from high latitudes of the primary star.

\subsection{An eclipsed flare in SV Cam}
To determine the geometry of the occulted flare,
we have considered the four contacts of the eclipse, identified 
at phases 1.024, 1.065, 1.118, 1.159 (Fig.~\ref{fig:eclipsezoom}) in
the light curve with 300~s binning. 
To simplify the problem, we assume that the
emitting region has a spherical shape, and we characterize the
flare with four variables: latitude ($\theta$), longitude
($\lambda$, with origin 0 the meridian of the each star that
is in front of the observer at $\phi$=0), height over the center 
of the hosting
star ($h$) and the size of the spherical emitting region ($R_3$). We
have made a grid of values for the 4 variables, and we then calculated
the times when the four contacts of the eclipse take place for each set of
values. We considered as valid results those that agree within 300~s
(1 bin) with the measured
times. There are four possible scenarios for the eclipse to happen:
\begin{enumerate}

\item primary star hosting the flare which is eclipsed by the
secondary,

\item self-eclipse in the secondary star,

\item self-eclipse in the primary star, 

\item self-eclipse by the primary star actually starting (first
contact) when the second star hides the flare, and ending behind the primary. 

\end{enumerate}

\begin{figure*}
   \centering
   \includegraphics[width=0.95\textwidth]{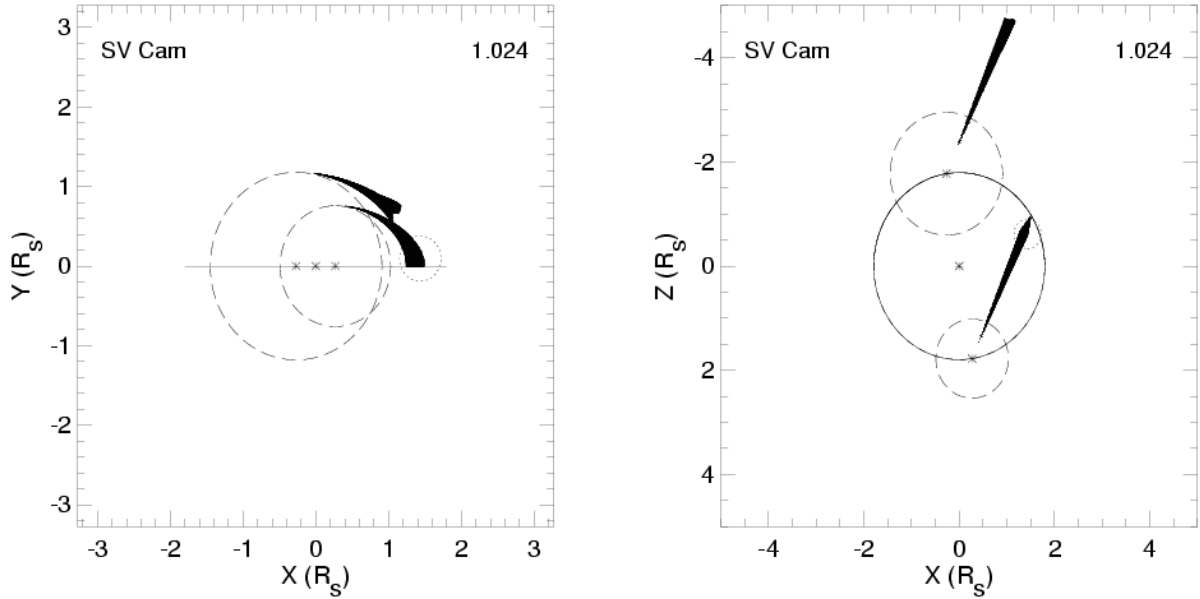}
   \caption{Relative positions of the two stars (dashed lines) during
   first contact ($\phi$=1.0240), as seen from the front (left
   panel) and top 
   (right panel). The solid line marks the orbit of the stars around
   the center of mass, while
   three asterisks mark the central positions of the two stars at
   first contact, and the center of the
   orbit. All the possible results are represented
   with black points indicating the center of the emitting region
   (only the solutions in the northern hemisphere are represented). An
   example of the eclipsed region with a size of $R_3$=0.28~$R_\odot$
   is shown in dotted line.}
   \label{fig:eclipse}
\end{figure*}

Possible solutions are plotted in Fig.~\ref{fig:eclipse} and
listed in Table~\ref{tab:eclipse}, and we will refer to them as cases
1, 2, 3 and 4 according to this Table.

We make use of the orbital data calculated by \citet{leh02}:
$R_1$, $R_2$=1.18, 0.76~R$_\odot$, $a_{12}$=3.60~R$_\odot$
(axis of the orbit), $i$=89.6$\degr$ (we can safely assume
$i$=90 to simplify equations), $M_1$, $M_2$=1.09, 0.70~M$_\odot$ and
circular orbits.
The solutions found in the first two cases (flare hosted by primary or
secondary, but eclipsed by the secondary) correspond actually to the same
points in the 3-D space. The equations corresponding to the first case
(primary star hosting a flare eclipsed by secondary) are:

\begin{displaymath}
x_1=-\frac{1}{2}\; a \sin(\phi), \quad y_1=0, \quad z_1=-\frac{1}{2}\; a \cos{\phi}
\end{displaymath}
\begin{displaymath}
x_2=-x_1, \quad y_2=0, \quad z_2=-z_1
\end{displaymath}
\begin{displaymath}
x_3=x_1+h \cos(\theta) \sin(\phi+\lambda), \quad 
y_3=y_1+h \sin(\theta), \quad  
\end{displaymath}
\begin{displaymath}
z_3=z_1+h \cos(\theta) \cos(\phi+\lambda)
\end{displaymath}

where $x$, $y$ correspond to the plane in front of the observer
(Fig.~\ref{fig:eclipse}), and $z$ is perpendicular to this plane
(positive towards the observer). $x_3$, $y_3$, $z_3$ correspond to the
position of the emitting region of the flare. 
In the first and fourth contacts the distance between the center
of the flare and the center of secondary star must be:

\begin{displaymath}
\Delta_1=\sqrt{(R_2+R_3)^2-(y_3-y_2)^2}
\end{displaymath}
and similarly, in the second and third contacts:
\begin{displaymath}
\Delta_2=\sqrt{(R_2-R_3)^2-(y_3-y_2)^2}
\end{displaymath}

\begin{figure*}
   \centering
   \includegraphics[width=0.9\textwidth]{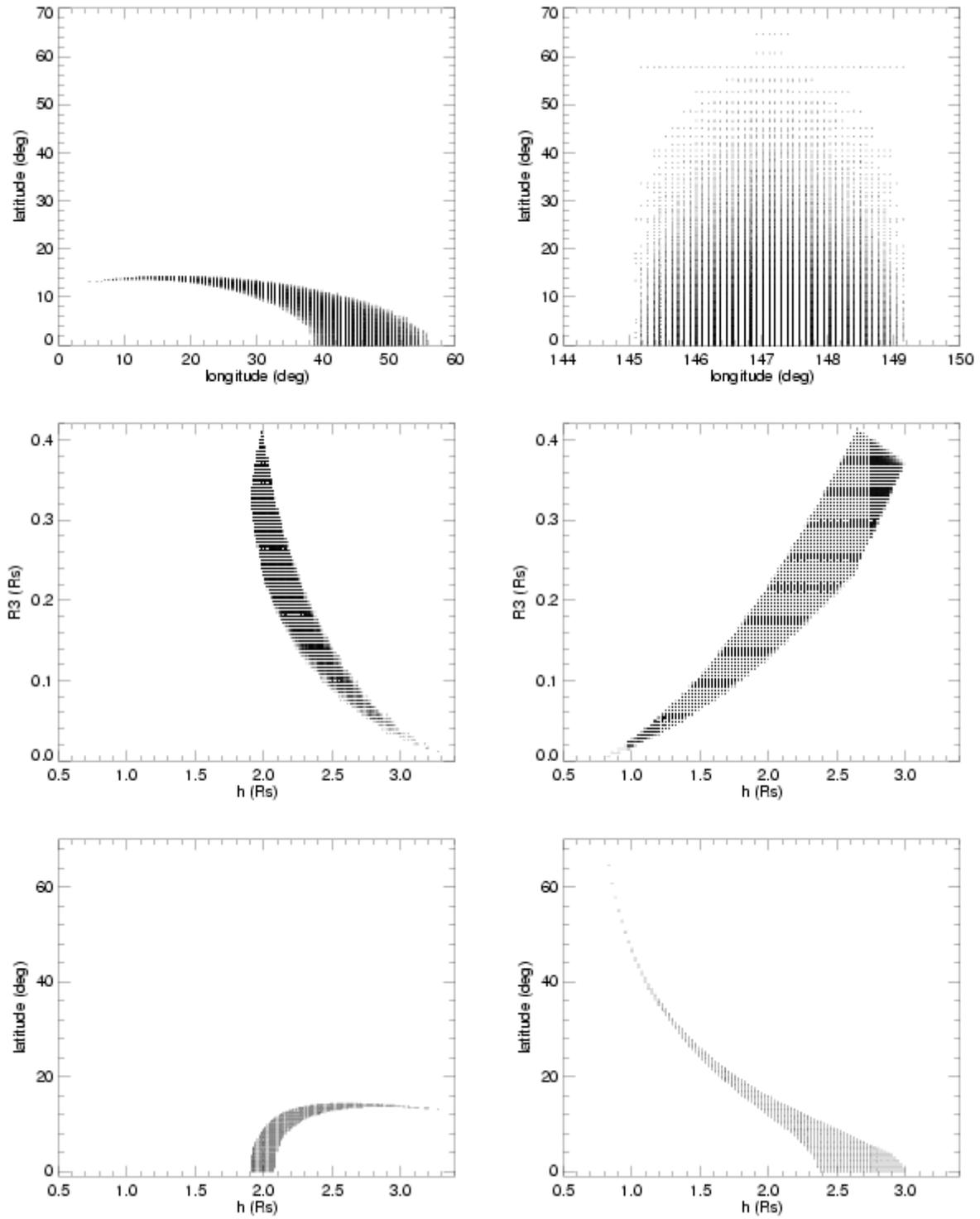}
   \caption{{\em Left:} Results in case 1 (primary flaring, secondary
   eclipsing). {\em Right:} Results in case 2 (self-eclipse in secondary).}
   \label{case12res}
\end{figure*}

\begin{figure*}
   \centering
   \includegraphics[width=0.9\textwidth]{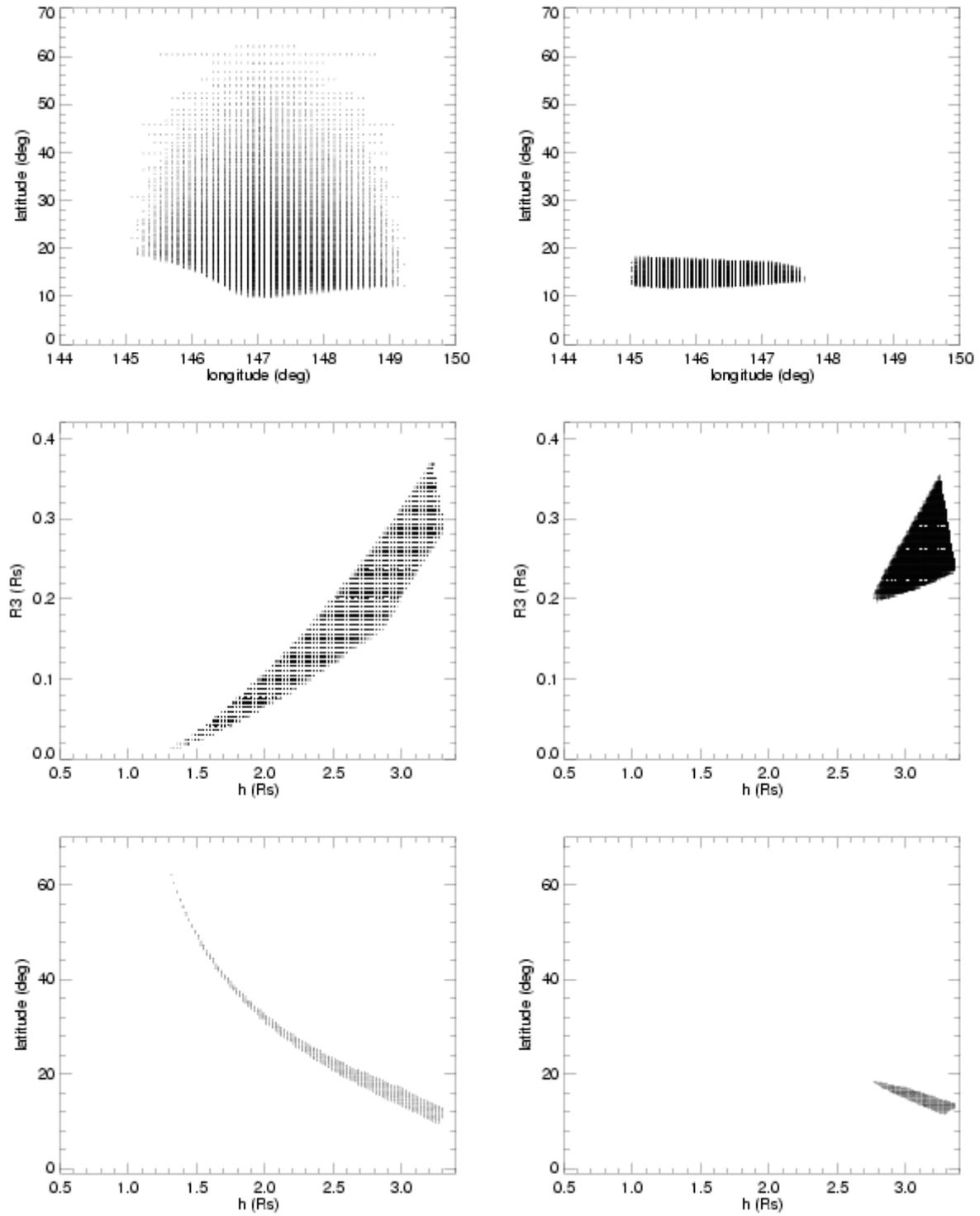}
   \caption{{\em Left:} Results in case 3 (self-eclipse entirely in
   primary). {\em Right:} Results in case 4 (primary flaring, eclipse
   starts in secondary, ends in primary).}
   \label{case34res}
\end{figure*}

Thus, in the four positions we will have four equations: $x_3=x_2+\Delta_1,
\quad  x_3=x_2+\Delta_2 \quad x_3=x_2-\Delta_2 \quad x_3=x_2-\Delta_1$
respectively, that we can use to calculate numerically 
the times of the four contacts for each point of the grid.
Some geometrical constraints can be added to ensure the totality of
the eclipse ($y_3 + R_3 \le y_2 + R_2$) and a valid position in $z$
($z_3 +\sqrt{(R_2+R_3)^2-(y_3-y_2)^2} < z_2$). 
Similar equations can be easily derived for the other cases.
An additional constraint was imposed in the cases 3 and 4, the only
scenarios in which $h$ reaches values larger than $a$: the centrifugal
force must be lower than the resulting gravitational forces of the two
stars. In this way, 
$h$+$R_3$ must be lower than $\sim$3.6~R$_\odot$.
The results found are shown in Table~\ref{tab:eclipse} and in
Figs.~\ref{case12res} and \ref{case34res}\footnote{These figures
are available in electronic version only.}. The trends  
relating the different variables are: in case 1, for an
increasing $h$, $R_3$ and $\lambda$ decrease, and $\theta$
increases; for the other 3 cases an increase in $h$ yields an increase in
$R_3$ and a decrease in $\theta$, with no trend for $\lambda$. 

\begin{table*}
\caption{Range of possible solutions (northern hemisphere only) 
of the eclipse of a flare in 2003.}\label{tab:eclipse} 
\begin{center}
\begin{footnotesize}
 \begin{tabular}{cccccccc}
\hline \hline
{Flaring star} & {Ecl. star} & $\theta$($\degr$) &$\lambda$($\degr$)  & 
$h$ (R$_\odot$) & $R_3$ (R$_\odot$) & $\log n_{\rm e}$ (cm$^{-3}$) & $B$ (G)\\
\hline
Pri & Sec & 0--14 &  4.7--55.7   & 1.91--3.28 & 0.012--0.41 &
10.6--12.9 & 66--930\\ 
Sec & Sec & 0--65 & 145.1--149.1 & 0.83--2.99 & 0.006--0.41 &
10.6--13.4 & 66--1600\\ 
Pri & Pri & 9.7--62  & 145.1--149.1 & 1.31--3.30 & 0.013--0.37 &
10.7--12.8 & 72--880 \\ 
Pri & S+P & 12--18 & 145.0--147.6 & 2.77--3.36 & 0.20--0.36 &
10.7--11.1 & 74--110 \\ 
\multicolumn{2}{c}{All~results} & 0--65 & $\sim$5--56 or $\sim$146 & 0.83--3.36 &
0.006--0.41 & 10.6--13.4 & 66--1600 \\ 
\hline
\end{tabular}
\end{footnotesize}
\end{center}

\end{table*}

We can also estimate the electron density ($n_{\rm e}$) of the emitting
region by calculating the Emission Measure ({\em EM})
before ($\phi \la$0.6) and during the flare prior to the eclipse 
($\phi \sim 0.87-1.02$) through a fit of the MOS spectra. A fit using 3
temperatures resulted in an average value of $\log T$ (K)=7.17 and
$\log$~{\em EM} (cm$^{-3}$)=53.17 in quiescence and $\log T$ (K)=7.22
and $\log$~{\em EM} (cm$^{-3}$)=53.38 during the flare.  
Since we can define {\it EM}$\sim 0.8 \, n_{\rm e}^2 \, V$, it is
possible to get the density $n_{\rm e}$ from the net flare {\it EM}, 
using $V=4 \pi/3 R_3^3$ (see
Table~\ref{tab:eclipse}). If we consider that the magnetic pressure, 
$B^2/(8 \pi)$, should be at least as large as the electron pressure
($2 n_e k T$), we can get the minimum magnetic field necessary
to have a stable structure in the flare.

The analysis of the eclipse leads to several interesting
conclusions: (i) the flare takes place at regions with $\theta
\la 65 \degr$; (ii) the emitting region is compact in size ($R_3 \la
0.41$~R$_\odot$), implying values of the electron density consistent
with those calculated from the {\em Chandra} and XMM-{\em Newton} 
spectra of active stars
at different temperatures \citep{sanz03,nes04,tes04}; 
(iii) the emitting region is
not necessarily attached to the surface of the star (the minimum $h$
found -- in case 2 -- implies also the minimum size and the highest
values of density) and therefore the loop of the flare likely 
has its emission concentrated in the apex.

\subsection{Eclipses and flares}
There are a few cases in the literature where the flares have been
eclipsed. Such an event is very useful to constrain some of the
characteristics of the flaring region. The technique most commonly used 
derives the position of the flare from the whole eclipse light curve.
This was the case of two flares in
the secondary star of Algol (B8V/K2III): \citet{schm99} using
BeppoSAX, observed a flare in a
polar region (there was no rotational modulation), with a loop height/size
$\la$0.6~R$_*$ (2~R$_\odot$), resulting in densities of at least
$\sim$10$^{11}$ cm$^{-3}$ and $B\ga 500$~G; \citet{sch03} using a more
detailed analysis, studied another flare in Algol observed with
XMM-{\it Newton}, resulting in a
nearly equatorial location with loop
heights $\sim$0.1~R$_*$ (0.3~R$_\odot$) and densities of several
times 10$^{11}$ cm$^{-3}$, resulting again in $B\sim 400$~G as the
minimum to accommodate the electron pressure found. More recently,
\citet{mit05} used the oscillations observed in a flare of the dM star
AT~Mic to constrain the flare loop length to $\sim$0.36~R$_\odot$ and
$B\sim 100$~G.
The first case reported in the literature was an analysis of an ASCA light
curve with lower statistics of VW~Cep \citep{cho98}. The authors use rather
qualitative assessments, and the ingress and egress times of the eclipse,
to conclude that the flare should be occuring near the pole, at a
height/size of 0.5~R$_*$ ($\sim$0.5~R$_\odot$) and $B\sim 80$~G.

\begin{figure}
   \centering
   \includegraphics[width=0.49\textwidth]{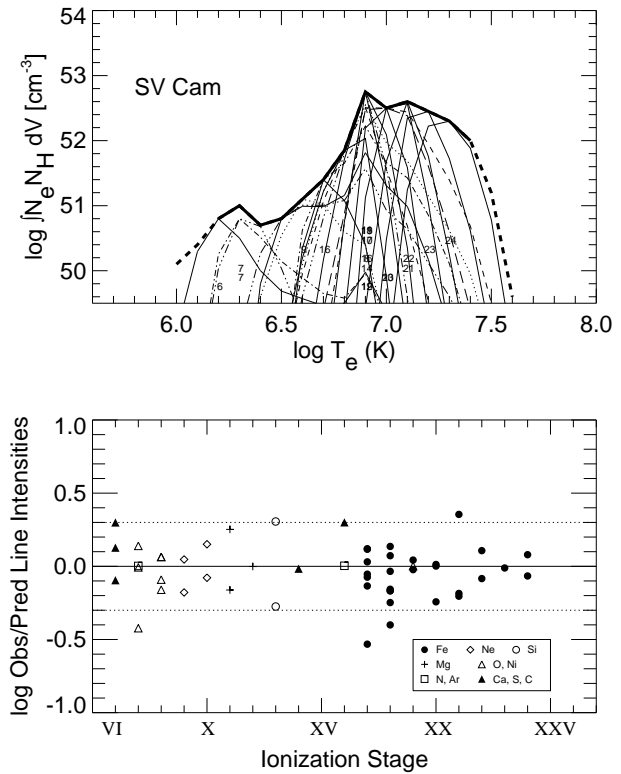}
   \caption{Emission Measure Distribution of SV Cam in Oct. 2003. Thin lines
   represent the relative contribution function for each ion (the
   emissivity function weighted by the EMD at each point). Small
   numbers indicate the ionization stages of the species. The observed
   to predicted line flux ratios for the ion stages in the upper
   figure are plotted in the lower figure. Dotted lines denote a
   factor of 2.}
   \label{emd}
\end{figure}

Although the techniques employed do not result in an exact
determination of the geometry of the loop, all these studies
need for a large (at least 100~G) 
magnetic field in order to explain
the presence of loops that have a relatively compact sizes. 
We will apply the technique used for SV~Cam to the analysis of
these flares in a future publication (Sanz-Forcada et al., in
preparation).

\subsection{Thermal structure and abundances}

The analysis of the thermal structure reveals an EMD (Fig.~\ref{emd},
Table~\ref{tab:emd})
very similar to 
the other active stars \citep{survey}, in which the EMD is 
dominated by a peak of material at
$\log T$(K)$\sim$6.9 with strong dependency on the rotational period of
the star, and with a large amount of material at higher
temperatures in the case of SV Cam 
that is also supported by the detection in the EPIC-PN
spectrum of the \ion{Fe}{xxv} complex at 6.7~keV.
In this case the
lack of statistics prevents the measurement of the \ion{O}{vii}
He-like triplet that would be useful to determine the electron density
at $\log T$(K)$\sim$6.3.

\begin{figure}
   \centering
   \includegraphics[width=0.45\textwidth]{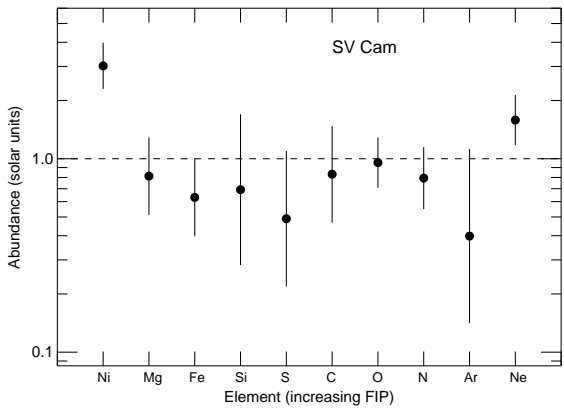}
   \caption{SV Cam coronal abundances in Oct. 2003. 
	Dashed line indicates solar photospheric abundances by \citet{anders}.}
   \label{abundances}
\end{figure}

Coronal abundances (Table~\ref{tab:abund},
Fig.~\ref{abundances}) are not as metal-deficient relative to the Sun
as they are for
other active stars 
\citep{huen01,aud03,sanz03,sanz04}, although a 
[Ne/Fe]=0.40$\pm$0.13 is lower 
than for other
stars of its activity level  \citep[e.g.,][]{dra01}. If we consider the 
uncertainties, most values are consistent with solar photosperic
abundances. Most of the uncertainty
comes from the determination of the Fe abundance, given the
difficulties in setting the continuum in the RGS spectrum. The rest of the
abundances were all calculated relative to Fe, therefore any shift in
its value will affect the others.
The lack of 
information on the photospheric abundances of SV Cam prevents us from drawing
any conclusions on the changes in abundance pattern taking place in
the corona, as it has been shown in \citet{sanz04}. 

\begin{table}
\caption{Emission Measure Distribution of SV Cam in Oct. 2003.}\label{tab:emd}
\begin{center}
\begin{small}
\begin{tabular}{cc}
\hline \hline
{log~$T$ (K)} & {log $\int N_{\rm e} N_{\rm H} {\rm d}V$
  (cm$^{-3}$)$^a$} \\
\hline
6.0 & 51.50: \\
6.1 & 50.40: \\
6.2 & 50.80$^{+0.20}_{-0.40}$  \\
6.3 & 51.00$^{+0.10}_{-0.30}$  \\
6.4 & 50.70$^{+0.20}_{-0.40}$  \\
6.5 & 50.80$^{+0.20}_{-0.40}$  \\
6.6 & 51.10$^{+0.20}_{-0.40}$  \\
6.7 & 51.40$^{+0.30}_{-0.30}$  \\
6.8 & 51.85$^{+0.25}_{-0.25}$  \\
6.9 & 52.75$^{+0.00}_{-0.00}$  \\
7.0 & 52.50$^{+0.20}_{-0.20}$  \\
7.1 & 52.60$^{+0.20}_{-0.20}$  \\
7.2 & 52.45$^{+0.15}_{-0.25}$  \\
7.3 & 52.30$^{+0.20}_{-0.30}$  \\
7.4 & 52.00$^{+0.30}_{-0.30}$  \\
7.5 & 51.20: \\
7.6 & 49.60: \\
\hline
\end{tabular}
\end{small}
\end{center}
$^a$Emission Measure, where $N_{\rm e}$ 
and $N_{\rm H}$ are electron and hydrogen densities, in
cm$^{-3}$. Error bars provided are not independent
between the different temperatures, see text.
\end{table}

\section{Conclusions}

 The corona of SV Cam has shown a very similar thermal
pattern to all the active stars, and although coronal abundances were
calculated, no conclusion can be drawn due to the lack of information
on the photospheric abundances. The most interesting information on
SV~Cam comes from its light curves.
The SV Cam corona was shown to be variable, with a change in flux
of at least $\sim$\,50\% in 36 months, and several flares. 
No coronal eclipses have been unequivocally identified during quiescence,
but the eclipse of a flare allowed us to constrain the size and
location of the emitting region of the flare. We found that the flare
is compact ($R_3 \sim 0.01-0.4~{\rm R}_\odot$), thus implying a range of
electron densities that 
is consistent with the values calculated from line ratios at different
temperatures for active stars.
While the quiescent emission of the
corona seems to be extended or concentrated at high latitudes, the
eclipsed flare has $\theta \la 65 \degr$,
in contrast with the polar flare observed in Algol by \citet{schm99}.

\begin{table}
\caption{Coronal abundances of the elements ([X/H], solar
  units) in SV Cam.}\label{tab:abund} 
\begin{center}
\begin{footnotesize}
 \begin{tabular}{lrcr}
\hline \hline
{X} & {FIP} & Reference$^a$ & [X/H] \\ 
    &  eV   & solar value & SV Cam  \\
\hline
 Ni &  7.63 & 6.25 & 0.48$\pm$0.12  \\
 Mg &  7.64 & 7.58 & $-$0.09$\pm$0.20  \\
 Fe &  7.87 & 7.67 & $-$0.20$\pm$0.20 \\
 Si &  8.15 & 7.55 & $-$0.16$\pm$0.39 \\
  S & 10.36 & 7.21 & $-$0.31$\pm$0.35 \\
  C & 11.26 & 8.56 & $-$0.08$\pm$0.25 \\
  O & 13.61 & 8.93 & $-$0.02$\pm$0.13 \\
  N & 14.53 & 8.05 & $-$0.10$\pm$0.16 \\
 Ar & 15.76 & 6.56 & $-$0.40$\pm$0.45 \\
 Ne & 21.56 & 8.09 & 0.20$\pm$0.13 \\
\hline
\end{tabular}
\end{footnotesize}
\end{center}
$^a$ Solar photospheric abundances from \citet{anders}, adopted in
this work, are expressed in logarithmic scale. 
Note that several values have been
updated in the literature, most notably the cases of Fe
\citep[now 7.45,][]{asp00}, O \citep[now
  $\sim$8.7,][]{all01,hol01} and C \citep[now 8.39,][]{all02}.

\end{table}

\begin{acknowledgements}
This research is based on observations obtained with XMM-Newton, an
ESA science mission with instruments and contributions directly funded
by ESA Member States and NASA. This research has made use
of NASA's Astrophysics Data System Abstract Service.
JS thanks the ESA Research Fellowship Program for support. 
We are grateful to the anonymous referee for the careful reading and useful
comments.
\end{acknowledgements}

\Online
\begin{table*}
\caption{XMM/RGS line fluxes of SV Cam$^a$}\label{tab:fluxes}
\tabcolsep 3.pt
\begin{scriptsize}
\begin{tabular}{lrcrrrl}
\hline \hline
 Ion & {$\lambda$$_{\mathrm {model}}$} &  
 log $T_{\mathrm {max}}$ & $F_{\mathrm {obs}}$ & S/N & ratio & Blends \\ 
\hline
\ion{Si}{xiii} &  6.6479 & 7.0 & 1.84e-14 &   3.9 & -0.27 & \ion{Si}{xiii}  6.6882 \\
\ion{Si}{xiii} &  6.7403 & 7.0 & 3.00e-14 &   3.7 &  0.31 & \ion{Mg}{xii}  6.7378, \ion{Si}{xiii}  6.7432 \\
\ion{Mg}{xii} &  8.4192 & 7.0 & 3.48e-14 &   5.3 & -0.00 & \ion{Mg}{xii}  8.4246 \\
\ion{Mg}{xi} &  9.1687 & 6.8 & 1.42e-14 &   5.4 & -0.16 &  \\
\ion{Mg}{xi} &  9.2312 & 6.8 & 3.31e-14 &   8.4 &  0.25 & \ion{Fe}{xxi}  9.1944, \ion{Ni}{xix}  9.2540, \ion{Fe}{xxii}  9.2630, \ion{Mg}{xi}  9.3143, \ion{Ni}{xxv}  9.3400 \\
\ion{Fe}{xxi} &  9.4797 & 7.0 & 2.23e-14 &   7.1 &  0.35 & \ion{Ne}{x}  9.4807, 9.4809 \\
No id. & 10.0200 &  & 2.03e-14 &   5.1 & \ldots & \ion{Na}{xi} 10.0232, 10.0286 \\
\ion{Ne}{x} & 10.2385 & 6.8 & 1.82e-14 &   7.0 & -0.08 & \ion{Ne}{x} 10.2396 \\
\ion{Fe}{xxiv} & 10.6190 & 7.3 & 2.84e-14 &   6.7 &  0.08 & \ion{Fe}{xix} 10.6414, 10.6491, 10.6840, \ion{Fe}{xvii} 10.6570, \ion{Fe}{xxiv} 10.6630 \\
\ion{Fe}{xvii} & 10.7700 & 6.8 & 6.05e-15 &   3.1 & -0.53 & \ion{Ni}{xxiii} 10.7214, 10.8491 , \ion{Fe}{xix} 10.8160\\
\ion{Fe}{xxiii} & 11.0190 & 7.2 & 3.16e-14 &   7.3 & -0.01 & \ion{Fe}{xxiii} 10.9810, \ion{Fe}{xxiv} 11.0290 \\
\ion{Fe}{xxiv} & 11.1760 & 7.3 & 1.47e-14 &   5.1 & -0.07 & \ion{Fe}{xvii} 11.1310, \ion{Fe}{xxiv} 11.1870 \\
\ion{Fe}{xvii} & 11.2540 & 6.8 & 1.96e-14 &   5.9 &  0.12 & \ion{Ni}{xxii} 11.2118, \ion{Fe}{xxiv} 11.2680, \ion{Fe}{xxiii} 11.2850, \ion{Ni}{xxi} 11.2908 \\
\ion{Fe}{xviii} & 11.4230 & 6.9 & 2.41e-14 &   6.5 &  0.07 & \ion{Fe}{xxii} 11.4270, \ion{Fe}{xxiv} 11.4320, \ion{Fe}{xxiii} 11.4580 \\
\ion{Fe}{xviii} & 11.5270 & 6.9 & 8.50e-15 &   3.8 & -0.40 & \ion{Fe}{xviii} 11.5270, \ion{Fe}{xxii} 11.4900, \ion{Ni}{xix} 11.5390, \ion{Ni}{xxi} 11.5390, \ion{Ne}{ix} 11.5440 \\
\ion{Fe}{xxii} & 11.7700 & 7.1 & 5.77e-14 &  10.5 & -0.08 & \ion{Fe}{xxiii} 11.7360, \ion{Ni}{xx} 11.8320, 11.8460 \\
\ion{Fe}{xxii} & 11.9770 & 7.1 & 1.67e-14 &   5.7 &  0.11 & \ion{Fe}{xxii} 11.9320, \ion{Fe}{xxi} 11.9750 \\
\ion{Ne}{x} & 12.1320 & 6.8 & 1.24e-13 &  15.8 &  0.15 & \ion{Ne}{x} 12.1321 \\
\ion{Fe}{xxi} & 12.2840 & 7.0 & 3.47e-14 &   8.4 & -0.20 & \ion{Fe}{xvii} 12.2660 \\
\ion{Ni}{xix} & 12.4350 & 6.9 & 3.23e-14 &   8.2 &  0.00 & \\
\ion{Fe}{xx} & 12.8240 & 7.0 & 4.23e-14 &   9.4 & -0.24 & \ion{Fe}{xxi} 12.8220, \ion{Fe}{xx} 12.8460, 12.8640 \\
\ion{Fe}{xx} & 12.9650 & 7.0 & 3.63e-14 &   8.4 &  0.01 & \ion{Fe}{xx} 12.9120, 12.9920, 13.0240 , \ion{Fe}{xix} 12.9330, 13.0220, \ion{Fe}{xxii} 12.9530\\
\ion{Fe}{xx} & 13.1530 & 7.0 & 3.13e-14 &   7.7 &  0.00 & \ion{Fe}{xx} 13.1370, 13.2740, \ion{Fe}{xxii} 13.2360, \ion{Fe}{xxi} 13.2487, \ion{Ni}{xx} 13.3090 \\
\ion{Ne}{ix} & 13.4473 & 6.6 & 3.46e-14 &   8.8 & -0.18 & \ion{Fe}{xx} 13.3850, \ion{Fe}{xix} 13.4230, 13.4620 \\
\ion{Fe}{xix} & 13.5180 & 6.9 & 5.84e-14 &  11.5 & -0.02 & \ion{Fe}{xix} 13.4970, \ion{Fe}{xxi} 13.5070, \ion{Ne}{ix} 13.5531 \\
\ion{Ne}{ix} & 13.6990 & 6.6 & 3.03e-14 &   8.4 &  0.05 & \ion{Fe}{xix} 13.6450, 13.6752, 13.7315, 13.7458 \\
\ion{Fe}{xix} & 13.7950 & 6.9 & 4.04e-14 &  12.9 &  0.04 & \ion{Fe}{xx} 13.7670, \ion{Ni}{xix} 13.7790, \ion{Fe}{xvii} 13.8250 \\
\ion{Fe}{xxi} & 14.0080 & 7.0 & 2.60e-14 &   7.8 & -0.19 & \ion{Ni}{xix} 14.0430, 14.0770 \\
\ion{Fe}{xviii} & 14.2080 & 6.9 & 4.72e-14 &  11.9 & -0.17 & \ion{Fe}{xviii} 14.2560, \ion{Fe}{xx} 14.2670 \\
\ion{Fe}{xviii} & 14.3730 & 6.9 & 2.47e-14 &   6.6 & -0.17 & \ion{Fe}{xx} 14.3318, 14.4207, 14.4600, \ion{Fe}{xviii} 14.3430, 14.4250, 14.4392 \\
\ion{Fe}{xviii} & 14.5340 & 6.9 & 1.10e-14 &   5.3 & -0.25 & \ion{Fe}{xviii} 14.4856, 14.5056, 14.5710, 14.6011 \\
\ion{O }{viii} & 14.8205 & 6.5 & 8.96e-15 &   6.6 &  0.06 & \ion{Fe}{xviii} 14.7820, 14.7867, \ion{O }{viii} 14.8207, \ion{Fe}{xx} 14.8276 \\
\ion{Fe}{xvii} & 15.0140 & 6.7 & 8.14e-14 &  19.8 & -0.13 & \ion{Fe}{xix} 15.0790 \\
\ion{O }{viii} & 15.1760 & 6.5 & 2.23e-14 &   9.9 &  0.06 & \ion{O }{viii} 15.1765, \ion{Fe}{xix} 15.1980 \\
\ion{Fe}{xvii} & 15.2610 & 6.7 & 3.15e-14 &  11.6 &  0.12 &  \\
\ion{Fe}{xvii} & 15.4530 & 6.7 & 6.95e-15 &   4.9 &  0.03 & \ion{Fe}{xix} 15.4136, \ion{Fe}{xviii} 15.4940, 15.5199, \ion{Fe}{xx} 15.5170 \\
\ion{Fe}{xviii} & 15.6250 & 6.8 & 1.63e-14 &   9.0 &  0.14 &  \\
\ion{Fe}{xviii} & 15.8700 & 6.8 & 7.83e-15 &   6.3 & -0.16 & \ion{Fe}{xviii} 15.8240 \\
\ion{O }{viii} & 16.0066 & 6.5 & 6.63e-14 &  18.3 & -0.09 & \ion{Fe}{xviii} 16.0040, 16.0710, 16.1590, \ion{O }{viii} 16.0055, \ion{Fe}{xix} 16.1100 \\
\ion{Fe}{xvii} & 16.7800 & 6.7 & 3.19e-14 &  12.8 & -0.05 &  \\
\ion{Fe}{xvii} & 17.0510 & 6.7 & 5.85e-14 &  14.9 & -0.07 & \ion{Fe}{xvii} 17.0960 \\
\ion{Fe}{xviii} & 17.6230 & 6.8 & 9.60e-15 &   7.0 & -0.03 &  \\
\ion{O }{vii} & 18.6270 & 6.3 & 3.19e-15 &   2.9 & -0.42 & \ion{Ca}{xviii} 18.6910 \\
\ion{O }{viii} & 18.9671 & 6.5 & 1.18e-13 &  17.3 & -0.16 & \ion{O }{viii} 18.9725 \\
\ion{Ca}{xvi} & 21.4500 & 6.7 & 4.14e-15 &   3.3 &  0.30 & \ion{Ca}{xvi} 21.4410 \\
\ion{O }{vii} & 21.6015 & 6.3 & 2.32e-14 &   7.8 &  0.14 & \ion{Ca}{xvi} 21.6100 \\
\ion{O }{vii} & 22.0977 & 6.3 & 1.14e-14 &   5.4 & -0.01 & \ion{Ca}{xvii} 22.1140 \\
\ion{N }{vii} & 24.7792 & 6.3 & 8.87e-15 &   6.1 &  0.00 & \ion{N }{vii} 24.7846 \\
\ion{Ar}{xvi} & 24.9910 & 6.7 & 8.61e-16 &   2.2 &  0.00 & \ion{Ar}{xvi} 25.0130, \ion{Ar}{xv} 25.0500 \\
\ion{C }{vi} & 26.9896 & 6.2 & 1.77e-15 &   3.2 &  0.30 & \ion{C }{vi} 26.9901 \\
\ion{C }{vi} & 28.4652 & 6.2 & 3.36e-15 &   2.9 &  0.13 & \ion{C }{vi} 28.4663 \\
\ion{S }{xiv} & 30.4270 & 6.5 & 1.79e-15 &   2.9 & -0.02 &  \\
\ion{C }{vi} & 33.7342 & 6.1 & 1.11e-14 &   7.0 & -0.09 & \ion{C }{vi} 33.7396 \\
\hline
\end{tabular}

{$^a$ Line fluxes (in erg cm$^{-2}$ s$^{-1}$) 
  measured in XMM/RGS SV Cam spectra. 
  log $T_{\rm max}$ indicates the maximum
  temperature (K) of formation of the line (unweighted by the
  EMD). ``Ratio'' is the log($F_{\mathrm {obs}}$/$F_{\mathrm {pred}}$) 
  of the line. 
  Blends amounting to more than 5\% of the total flux for each line are
  indicated.}
\end{scriptsize}
\end{table*}

\end{document}